\documentclass[runningheads]{llncs}
\usepackage{graphicx, float}
\usepackage{hyperref,amsfonts,amsmath,xcolor,nicefrac}

\usepackage{listings}

\definecolor{codegreen}{rgb}{0,0.6,0}
\definecolor{codegray}{rgb}{0.5,0.5,0.5}
\definecolor{codepurple}{rgb}{0.58,0,0.82}
\definecolor{backcolour}{rgb}{0.95,0.95,0.92}

\lstdefinestyle{mystyle}{
    backgroundcolor=\color{white},   
    commentstyle=\color{codegreen},
    keywordstyle=\color{magenta},
    numberstyle=\tiny\color{codegray},
    stringstyle=\color{codepurple},
    basicstyle=\ttfamily\scriptsize,
    breakatwhitespace=false,         
    breaklines=true,                 
    captionpos=b,                    
    keepspaces=true,                 
    numbers=left,                    
    numbersep=5pt,                  
    showspaces=false,                
    showstringspaces=false,
    showtabs=false,                  
    tabsize=2
}

\usepackage{xparse}
\NewDocumentCommand{\code}{v}{%
\texttt{\textcolor{black}{#1}}%
}

\usepackage{listings}

\begin{document}
\title{
On the design of Monte-Carlo particle coagulation solver interface: 
a CPU/GPU Super-Droplet Method case study with PySDM%
\thanks{Funding: Foundation for Polish Science (POIR.04.04.00-00-5E1C/18-00).}}
\titlerunning{Monte-Carlo particle coagulation: CPU/GPU case study with PySDM}
\author{Piotr Bartman\inst{1}\orcidID{0000-0003-0265-6428} \and
Sylwester Arabas\inst{1}\orcidID{0000-0003-2361-0082}}
\authorrunning{P. Bartman \& S. Arabas}
\institute{Jagiellonian University, Kraków, Poland\\~\\
\email{piotr.bartman@doctoral.uj.edu.pl, sylwester.arabas@uj.edu.pl}}
\maketitle 
\begin{abstract} 
Super-Droplet Method (SDM) is a probabilistic Monte-Carlo-type model of particle coagulation process, an alternative to the mean-field formulation of Smoluchowski.
SDM as an algorithm has linear computational complexity with respect to the state vector length, the state vector length is constant throughout simulation, and most of the algorithm steps are readily parallelizable.
This paper discusses the design and implementation of two number-crunching backends for SDM implemented in PySDM, a~new free and open-source Python package for simulating the dynamics of atmospheric aerosol, cloud and rain particles. 
The two backends share their application programming interface (API) but leverage distinct parallelism paradigms, target different hardware, and are built on top of different lower-level routine sets. First offers multi-threaded CPU computations and is based on Numba (using Numpy arrays). Second offers GPU computations and is built on top of ThrustRTC and CURandRTC (and does not use Numpy arrays). 
In the paper, the API is discussed focusing on: data dependencies across steps, parallelisation opportunities, CPU and GPU implementation nuances, and algorithm workflow. Example simulations suitable for validating implementations of the API are presented.
\end{abstract}

\section{Introduction}

The Super-Droplet Method (SDM) introduced in \cite{Shima_et_al_2009} is a computationally
  efficient Monte-Carlo type algorithm for modelling the process of collisional growth (coagulation)
  of particles.
SDM was introduced in the context of atmospheric modelling, in particular 
  for simulating the formation of cloud and rain 
  through particle-based simulations.
Such simulations couple a grid-based computational fluid-dynamics (CFD) core with a
  probabilistic, so-called super-particle (hence the algorithm name), 
  representation of the particulate phase, and constitute a tool
  for comprehensive modelling of aerosol-cloud-precipitation interactions (e.g.,~\cite{Andrejczuk_et_al_2010,Arabas_and_Shima_2013,Dziekan_et_al_2019,Hoffmann_et_al_2017}; see~\cite{Grabowski_et_al_2019} for a review).
The probabilistic character of SDM is embodied, among other aspects, in the assumption of each
  super-particle representing a multiple number of modelled droplets with the same attributes 
  (including particle physicochemical properties and position in space). 
The super-droplets are thus a coarse-grained view of droplets both in physical  and attribute space. 
    
The probabilistic description of collisional growth has a wide range of applications across
  different domains of computational sciences (e.g., astrophysics, aerosol/hydrosol technology including 
  combustion).
While focused on SDM and depicted with atmospheric phenomena examples, the material presented herein 
  is generally applicable in development of software implementing other Monte-Carlo type schemes for coagulation (e.g., Weighted Flow Algorithms \cite{DeVille_et_al_2011} and other, see
  Sect.~1 in \cite{Shima_et_al_2020}), particularly when sharing the concept of super-particles.

The original algorithm description \cite{Shima_et_al_2009}, the relevant patent applications (e.g., \cite{EP1847939A3}) and several subsequent works scrutinising 
  SDM (e.g., \cite{Arabas_et_al_2015,Dziekan_et_al_2017,Li_et_al_2017,Unterstrasser_et_al_2020} expounded upon the algorithm characteristics from users' (physicists') point of view.
There are several CFD packages implementing SDM including: 
SCALE-SDM \cite{Sato_et_al_2018} and UWLCM \cite{Dziekan_et_al_2019}; however the implementation aspects were not 
  within the scope of the works describing these development.
The aim of this work is to discuss the algorithm from software developer's perspective.
To this end, the scope of the discussion covers: data dependencies, parallelisation opportunities,
  state vector structure and helper variable requirements, minimal set of computational kernels 
  needed to be implemented and the overall algorithm workflow.
These concepts are depicted herein with pseudo-code-mimicking Python snippets
  (syntactically correct and runnable), but the solutions introduced are not bound to a
  particular language.
In contrast, the gist of the paper is the language-agnostic API (i.e., application programming interface)
  proposal put forward with the very aim of capturing the implementation-relevant nuances of the
  algorithm which are, arguably, tricky to discern from existing literature on SDM, yet which have 
  an impact on the simulation performance.
The API provides an indirection layer separating higher-level physically-relevant 
  concepts from lower-level computational kernels.
  
Validity of the proposed API design has been demonstrated with two distinct backend implementations
  included in a newly developed simulation package PySDM \cite{Bartman_et_al_2021}.
The two backends share the programming interface, while differing substantially in the
  underlying software components and the targeted hardware (CPU vs. GPU).
PySDM is free/libre and open source software, presented results are readily reproducible
  with examples shipped with PySDM (\url{https://github.com/atmos-cloud-sim-uj/PySDM}).

The remainder of this paper is structured as follows. 
Section~\ref{sec:background} briefly introduces the SDM algorithm  through a juxtaposition against the alternative classic Smoluchowski's coagulation equation (SCE).
Section~\ref{sec:API} covers the backend API.
Section~\ref{sec:examples} presents simulations performed with PySDM
  based on benchmark setups from literature documenting CPU and GPU performance of the implementation.
Section~\ref{sec:concl} concludes the work enumerating the key points brought out in the paper.

\section{SDM as compared to SCE}\label{sec:background}

\subsection{Mean-field approach: Smoluchowski's coagulation equation}

Population balance equation which describes collisional growth is historically known as the Smoluchowski's coagulation equation (SCE) and was introduced in \cite{Smoluchowski_1916b,Smoluchowski_1916a} (for a classic and recent overviews, see e.g. \cite{Chandrasekhar_1943} and \cite{Hansen_2018}, respectively). 
It~is formulated under the mean-field assumptions of sufficiently large well-mixed system and of neglected correlations between numbers of droplets of different sizes (for discussion in the context of SDM, see also \cite{Dziekan_et_al_2017}). 

Let function $c(x, t) : \mathbb{R}^+\times \mathbb{R}^+ \rightarrow \mathbb{R}^+$ correspond to particle size spectrum and describe the average concentration of particles with size defined by $x$ at time $t$ in a volume~$V$. 
Smoluchowski's coagulation equation describes evolution of the spectrum in time due to collisions. For convenience, $t$ is skipped in notation: $c(x) = c(x, t)$, while $\dot{c}$ denotes partial derivative with respect to time.
\begin{eqnarray}
    \dot{c}(x) = \frac{1}{2} \int_{0}^{x} a(y, x-y) c(y) c(x-y) dy - \int_{0}^{\infty} a(y, x) c(y) c(x) dy
\end{eqnarray}
where $a(x_1, x_2)$ is the so-called kernel which defines the rate of collisions (and~coagulation) between particles of sizes $x_1$ and $x_2$ and $a(x_1, x_2) = a(x_2, x_1)$. The first term on the right-hand side is the production of particles of size $x$ by coalescence of two smaller particles and the factor $\nicefrac{1}{2}$ is for avoiding double counting. The second term represents the reduction in number of colliding particles due to coagulation. 

The Smoluchowski's equation has an alternative form that is discrete in size space. Let $x_0$ be the smallest considered difference of size between particles, $x_i = i x_0$, $i \in \mathbb{N}$ and $c_i = c(x_i)$ then:
\begin{eqnarray}\label{eq:smol}
    \dot{c_i} = \frac{1}{2} \sum\limits_{k=1}^{i-1} a(x_k, x_{i-k}) c_k c_{i-k} - \sum\limits_{k=1}^{\infty} a(x_k, x_i) c_k c_i
\end{eqnarray}

Analytic solutions to the equation are known only for simple kernels \cite{Hansen_2018}, such as: constant $a(x_1, x_2) = 1$, additive $a(x_1, x_2) = x_1 + x_2$ (Golovin's kernel \cite{Golovin_1963}) or multiplicative $a(x_1, x_2) = x_1 x_2$. 
Taking atmospheric physics as an example, collisions of droplets within cloud occur by differentiated movements of particles caused by combination of gravitational, electrical, or aerodynamic forces, where gravitational effects dominate.
As such, sophisticated kernels are needed to describe these phenomena, and hence numerical methods are required for solving the coagulation problem. 
However, when multiple properties of particles (volume, chemical composition, etc.) need to be taken into account, the numerical methods for SCE suffer from the curse of dimensionality due to the need to distinguish particles of same size $x$ but different properties.

Additionally, it is worth to highlight that, in practice, the assumptions of the Smoluchowski equation may be difficult to meet.
First, the particle size changes at the same time due to processes other than coalescence (e.g., condensation/evaporation).
Second, it is assumed that the system is large enough and the droplets inside are uniformly distributed, which in turn is only true for a small volume in the atmosphere.
Moreover, using the Smoluchowski's equation that describes evolution of the mean state of the system, leads to deterministic simulations.
The alternative to Smoluchowski equation are Monte-Carlo methods based on stochastic model of the collision-coalescence process.
Stochastic approach enables simulating ensembles of realisations of the process aiding studies of rare (far-from-the-mean) phenomena like rapid precipitation onset (see \cite{Grabowski_et_al_2019}).

\subsection{Monte-Carlo approach: Super-Droplet Method (SDM)} \label{sec:SDM}

\begin{figure}[t]
    \centering
    \includegraphics[width=\textwidth]{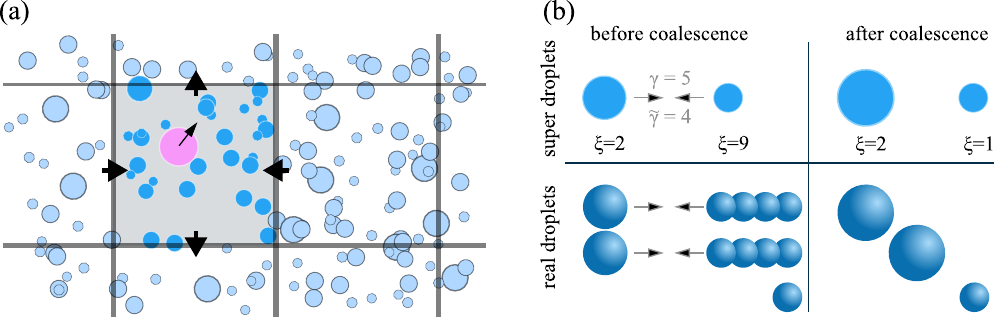}
    \caption{(a): Schematic of how a CFD grid is populated with immersed super-particles; (b) schematic of super-particle collision (based on Fig.1 in \cite{Shima_et_al_2009}), the parameters $\gamma$ and $\tilde{\gamma}$ are pertinent to representation of multiple collisions within a single coalescence event.}
    \label{fig:grid}
\end{figure}

The stochastic coalescence model for cloud droplet growth was already analysed by Gillespie in \cite{Gillespie_1972}. 
There remained however still challenges (elaborated on below) related to computational and memory complexity which are 
  addressed by SDM.
The original formulation of SDM given in \cite{Shima_et_al_2009} was presented in the context of liquid-phase
  atmospheric clouds, and extended to cover clouds featuring ice phase of water in \cite{Shima_et_al_2020}).

Tracking all particles in large simulations has immeasurable cost, which is why the notion of super-particles is introduced.
For convenience, let $\Omega$ be an indexed set of super-droplets for which collisions are considered (e.g., particles within one CFD grid cell, see Fig.~\ref{fig:grid}a with the blue droplets in a shaded cell depicting set $\Omega$).
Each element of this set has unique index $i \in [0, 1, ..., n(\Omega)-1]$, where $n(\Omega)$ is the number of considered super-droplets.
Attributes related to a specific super-droplet $i$ are denoted by $attr_{[i]}$. 
Note that only extensive attributes (volume, mass, number of moles) are considered within the coalescence scheme.

In SDM, each super-droplet represents a set of droplets with the same attributes (including position) and is assigned with a {\it multiplicity} denoted by $\xi \in \mathbb{N}$ and which can be different in each super-droplet. 
The multiplicity $\xi_{[i]}$ of super-droplet of size $x_{[i]}$ is conceptually related to $c_i=c(x_i)$ from equation~(\ref{eq:smol}). 
The number of super-droplets must be sufficient to cover the phase space of particle properties. The higher the number, the smaller the multiplicities and the higher fidelity of discretisation
(see \cite{Shima_et_al_2009,Shima_et_al_2020} for discussion).

Table~\ref{tab:smol_vs_sdm} summarises the main differences between the SCE and SDM.
The SDM algorithm steps can be summarised as follows.

\paragraph{\bf Step 1: cell-wise shuffling}~\\
In SCE, all considered droplets (e.g., those within a grid cell in a given time step) are compared with each other and a fraction of droplets of size $x_{[i]}$ collides with a fraction droplets of size $x_{[j]}$ in each time step. 
In contrast, in SDM a random set of $\lfloor n(\Omega)/2 \rfloor$ non-overlapping pairs is considered in each time step.

To ensure linear computational complexity, the mechanism for choosing random non-overlapping pairs must also have at most linear computational complexity. 
It can be obtained by permuting an indexed array of super-droplets where each pair consist of super-droplets with indices $(2j)$ and $(2j+1)$ where $j \in [0, 1, 2, \dots, \lfloor n(\Omega) / 2 \rfloor)$, see appendix A in \cite{Shima_et_al_2009}. 

\paragraph{\bf Step 2: collision probability evaluation}~\\
If a fraction of droplets is allowed to collide, a collision of two super-droplets can produce third super-droplet which has different size. 
To represent such collision event, a new super-droplet would need to be added to the system what leads to: increasing the number of super-droplets in the system, smaller and smaller multiplicities and increasing memory demands during simulation. 
Moreover, it is hard to predict at the beginning of a simulation how much memory is needed (see discussion in \cite{Jensen_et_al_2008}).
Adding new super-droplets can be avoided by mapping colliding super-droplets to a multi-dimensional grid (e.g., 2D for particle volume and its dry size), simulating coalescence on this grid using SCE-based approach and discretising the results back into Lagrangian particles as proposed in \cite{Andrejczuk_et_al_2010}.
Such approach however, not only entails additional computational cost, but most importantly is characterised by the very curse of dimensionality and numerical diffusion that were meant to be avoided by using particle-based technique.

SDM avoids the issue of unpredictable memory complexity stemming from the need of allocating space for freshly collided super-particle of size that differs from the size of the colliding ones.
To this end, in SDM only collisions of all of $\min\{\xi_{[i]},\xi_{[j]}\}$ droplets are considered, and thus collision of two super-droplets always produces one or two super-droplets (see Fig.~\ref{fig:grid}b). This means there is no need for adding extra super-droplets to the system during simulation.

For each pair, probability of collision $p$ is up-scaled by the number of real droplets represented by the super-droplet with higher multiplicity ($\max\{\xi_{[i]},\xi_{[j]}\}$). 
As a result, computational complexity is $\mathcal{O}(n(\Omega))$ instead of $\mathcal{O}(n(\Omega)^2)$.

The evaluated probability of collision of super-droplets 
requires further up-scaling due to the reduced amount of sampled
candidate pairs as outlined in Step~1 above.
To this end, the probability is multiplied by the ratio of
the number of all non-overlapping pairs $(n(\Omega)^2 - n(\Omega))/2$ to the number of considered candidate pairs $\lfloor n(\Omega)/2 \rfloor$. 

For each selected pair of $j$-th and $k$-th super-droplets, the probability $p$ of collision of the super-droplets within time interval $\Delta t$ is thus evaluated as:
\begin{equation}
    p = a(v_{[j]}, v_{[k]}) \max\{\xi_{[j]} , \xi_{[k]}\} \frac{(n(\Omega)^2 - n(\Omega))/2}{\lfloor n(\Omega)/2 \rfloor} \frac{\Delta t}{\Delta V} 
    \label{eq:p}
\end{equation}
where $a$ is a coalescence kernel and $\Delta V$ is the considered volume assumed to be
mixed well enough to neglect the positions of particles when evaluating the probability of collision. 

\paragraph{\bf Step 3: collision triggering and attribute updates}~\\
In the spirit of Monte-Carlo methods, the collision events are triggered 
  with a random number $\phi_\gamma\sim Uniform[0, 1)$, and the 
  rate of collisions (per timestep) $\gamma = \left\lceil p - \phi_\gamma \right\rceil$
with $\gamma\in \mathbb{N}$.
Noting that the rate $\gamma$ can be greater than 1, further adjustment is introduced to represent multiple collision of super-particles:
\begin{equation}
    \tilde{\gamma} = \min\{\gamma, \lfloor\xi_{[j]} / \xi_{[k]}\rfloor\}
    \label{eq:gamma2}
\end{equation}
where $\xi_{[j]} \ge \xi_{[k]}$ was assumed (without losing generality).
The conceptual view of collision of two super-droplets is intuitively depicted in Fig.~\ref{fig:grid}b, presented example corresponds to the case of $\gamma=5$ and $\tilde{\gamma}=\min\{5, \lfloor 9/2 \rfloor\}=4$.

\begin{table}[t]
\caption{Comparison of SCE and SDM approaches to modelling collisional growth.}\label{tab:smol_vs_sdm}
\resizebox{\textwidth}{!}{
\begin{tabular}{@{}c|c|c@{}}
 & SCE (mean-field) & SDM (probabilistic)      \\ 
\hline
\multicolumn{1}{c|}{considered pairs} & \multicolumn{1}{c|}{all (i,j) pairs}         & \multicolumn{1}{c}{\begin{tabular}[c]{@{}c@{}}random set of $n(\Omega)/2$ non-overlapping pairs, \\
probability up-scaled by  $(n(\Omega)^2 - n(\Omega))/2$  to  $n(\Omega)/2$  ratio\end{tabular}} \\
\multicolumn{1}{c|}{comp. complexity} & \multicolumn{1}{c|}{$\mathcal{O}(n(\Omega)^2)$}         & \multicolumn{1}{c}{$\mathcal{O}(n(\Omega))$} \\ 
\multicolumn{1}{c|}{collisions}       & \multicolumn{1}{c|}{colliding a fraction of $\xi_{[i]}$, $\xi_{[j]}$}         & \multicolumn{1}{c}{collide all of min\{$\xi_{[i]}$, $\xi_{[j]}$\} (all or nothing)} \\
\multicolumn{1}{c|}{collisions triggered}       & \multicolumn{1}{c|}{every time step}         & \multicolumn{1}{c}{by comparing probability with a random number}
\end{tabular}
}
\end{table}

As pointed out in Step 2 above, SDM is formulated in a way assuring that
each collision event produces one or two super-droplets.
A collision event results in an update of super-droplet attributes denoted with $A \in \mathbb{R}^{n_{\text{attr}}}$ where $n_{\text{attr}}$ is the number of considered extensive particle properties.
During coalescence, particle positions remain unchanged, values of extensive attributes of collided droplets add up, while the multiplicities are either reduced or remain unchanged. 
This corresponds to the two considered scenarios defined in points (5) (a) and (b) in Sec.~5.1.3 in~\cite{Shima_et_al_2009} and when expressed using the symbols used herein, with attribute values after collision denoted by hat, gives: 
\begin{enumerate}
  \item $\xi_{[j]} - \tilde{\gamma}\xi_{[k]} > 0$
    \begin{equation}
      \begin{split} \label{eq:coal_attr_0}
        \hat\xi_{[j]} &= \xi_{[j]} - \tilde{\gamma}\xi_{[k]} \\
        \hat A_{[j]} &=  A_{[j]}
      \end{split}
    \quad\quad
      \begin{split}
        \hat\xi_{[k]} &= \xi_{[k]} \\
        \hat A_{[k]} &=  A_{[k]} + \tilde{\gamma} A_{[j]}
      \end{split}
    \end{equation}
  \item $\xi_{[j]} - \tilde{\gamma}\xi_{[k]} = 0$
    \begin{equation}
      \begin{split}
        \hat\xi_{[j]} &= \lfloor\xi_{[k]}/2\rfloor \\
         \hat A_{[j]} &= \hat A_{[k]}
      \end{split}
    \quad\quad
      \begin{split}
        \hat\xi_{[k]} &= \xi_{[k]} - \lfloor\xi_{[k]}/2\rfloor \\
        \hat A_{[k]} &=  A_{[k]} + \tilde{\gamma} A_{[j]}
      \end{split} 
      \label{eq:coal}
    \end{equation}
\end{enumerate}
Case 1 corresponds to the scenario depicted in Fig.~\ref{fig:grid}b.
In case 2, all droplets following a coalescence event have the same values of extensive attributes (e.g.,~volume), thus in principle could be represented with a single super droplet. 
However, in order not to reduce the number of super-droplets in the system, the resultant super droplet is split in two.  
Since integer values are used to represent multiplicities, in the case of $\xi_{[k]}=1$ in eq.~(\ref{eq:coal}), splitting is not possible and the $j$-th super-droplet is removed from the system.

\section{Backend API}\label{sec:API}

The proposed API is composed of four data structures (classes) and a set of library routines (computational kernels).
Description below outlines both the general, implementation-independent, structure of the API, as well as selected aspects pertaining to
the experience from implementing CPU and GPU backends in the PySDM
package.
These were built on top of the Numba \cite{Lam_et_al_2015} and ThrustRTC
  Python packages, respectively.
Numba is an just-in-time (JIT) compiler for Python code, it features
  extensive support for Numpy and
  features multi-threading constructs akin to the OpenMP infrastructure.
ThrustRTC uses the NVIDIA CUDA real-time compilation infrastructure
  offering high-level Python interface for execution of both built-in
  and custom computational kernels on GPU.

In multi-dimensional simulations coupled with a grid-based
  CFD fluid flow solver, the positions of droplets within
  the physical space are used to split the super-particle population
  among grid cells (CFD solver mesh).
Since positions of droplets change based on the fluid flow and
  droplet mass/shape characteristics, the particle-cell mapping
  changes throughout the simulation.
In PySDM, as it is common in cloud physics applications, collisions are considered only among particles belonging to the same 
  grid cell, and the varying number of particles within a cell is needed
  to be tracked at each timestep to evaluate the super-particle collision rates.
Figure~\ref{fig:grid}a outlines the setting.

\subsection{Data structures and simulation state}

The \code{Storage} class is a base container which is intended to
  adapt the interfaces of the underlying implementation-dependent 
  array containers (in PySDM: Numpy or ThrustRTC containers for CPU and
  GPU backends, respectively).
This make the API independent of the underlying storage layer.

The \code{Storage} class has 3 attributes: 
  data (in PySDM: an instance of Numpy \code{ndarray} or an instance of ThrustRTC \code{DVVector}), 
  shape (which specifies size and dimension) and 
  dtype (data type: \code{float}, \code{int} or \code{bool}).
The proposed API employs one- and two-dimensional arrays, implementations
  of the \code{Storage} class feature an indirection layer 
  handling the multi-dimensionality in case the underlying 
  library supports one-dimensional indexing only (as in ThrustRTC).
The two-dimensional arrays are used for representing multiple extensive attributes (with row-major memory layout).
In general, structure-of-arrays layout is used within PySDM.

\code{Storage} handles memory allocation and optionally the 
  host-device (CPU-accessible and GPU-accessible memory) data transfers.
Equipping \code{Storage} with an override of the \code{[ ]} operator
  as done in PySDM can be helpful for debugging and unit testing (in Python, \code{Storage} instances may then be directly used with \code{max()}, \code{min()}, etc.), care needs to be taken to ensure memory-view semantics for non-debug usage, though. 
Explicit allocation is used only (once per simulation),

The \code{IndexedStorage} subclass of \code{Storage} is intended as
  container for super-particle attributes.
In SDM, at each step of simulation a different order of particles needs
  to be considered. 
To avoid repeated permutations of the attribute values, the \code{Index}
  subclass of \code{Storage} is introduced. 
One instance of \code{Index} is shared between \code{IndexedStorage}
  instances and is referenced by the \code{idx} field.
  
The \code{Index} class features permutation and array-shrinking logic
  (allowing for removal of super-droplets from the system).
To permute particle order, it is enough to shuffle
  \code{Index}.
To support simulations in multiple physical dimensions, \code{Index} features
  sort-by-key logic where a cell id attribute is used as the key.

The \code{PairIndicator} class provides an abstraction layer facilitating
  pairwise operations inherent to several steps of the SDM workflow.
In principle, it represents a Boolean flag per each super-particle
  indicating weather in the current state (affected by random
  shuffling and physical displacement of particles), a given
  particle is considered as first-in-a-pair.
Updating \code{PairIndicator}, it must be ensured that the next
  particle according to a given \code{Index} is the second one in 
  a pair -- i.e., resides in the same grid cell.
The \code{PairIndicator} is also used to handle odd and even 
  counts of super-particles within a cell (see also Fig.~\ref{fig:pair_flag}).
The rationale to store the pair-indicator flags, besides potential speedup,
  is to separate the cell segmentation logic pertinent to the internals of 
  SDM implementation from potentially user-supplied kernel code.

Figure~\ref{lst:vars} lists a minimal set of instances of the
  data structures constituting an SDM-based simulation state.

\begin{figure}[t]
\begin{lstlisting}[language=Python]
idx = Index(N_SD, int)
multiplicities = IndexedStorage(idx, N_SD, int)
attributes = IndexedStorage(idx, (N_ATTR, N_SD), float)
volume_view = attributes[0:1, :]

cell_id = IndexedStorage(idx, N_SD, int)
cell_idx = Index(N_CELL, int)
cell_start = Storage(N_CELL + 1, int)

pair_prob = Storage(N_SD//2, float)
pair_flag = PairIndicator(N_SD, bool)
        
u01 = Storage(N_SD, float)
\end{lstlisting}
  \caption{Simulation state example with {N\_SD} super-particles, {N\_CELL} grid cells and {N\_ATTR} attributes.}\label{lst:vars}
\end{figure}

\subsection{Algorithm workflow and API routines}

The algorithm workflow, coded in Python using the proposed data 
  structures, and divided into the same algorithm steps as outlined
  in Sec.~\ref{sec:background} is given in Fig.~\ref{fig:algo}.
An additional Step~0 is introduced to account for handling
  the removal of zero-multiplicity super-particles at the
  beginning of each timestep.

The \code{cell_id} attribute represents particle-cell mapping.
Furthermore, the \code{cell_idx} helper \code{Index} instance can be used
  to specify the order in which grid cells are traversed -- 
  for parallel execution scheduling purposes.
Both \code{cell_id} and \code{cell_idx} are updated
  before entering into the presented block of code.

Step~1 logic begins with generation of random numbers used
  for random permutation of the particles (in PySDM, the CURandRTC
  package is used).
It serves as a mechanism for random selection of super-particle pairs
  in each grid cell.
First, a shuffle-per-cell step is done in which the \code{cell_start}
  array is updated to indicate the location within cell-sorted \code{idx}
  where sets of super-particles belonging to a given grid cell 
  start.
Second, the \code{pair_flag} indicator is updated using the
  particle-cell mapping embodied in \code{cell_start}.

In PySDM, the shuffle-per-cell operation is implemented with two alternative strategies depending on the choice of the backend.
A parallel sort with a random key is used on GPU, while 
the CPU backend uses a serial $\mathcal{O}(n)$ permutation algorithm (as in Appendix A in \cite{Shima_et_al_2009}).
This choice was found to offer shortest respective execution times for the considered setup.
In general, the number of available threads and the number of droplets and cells considered will determine optimal choice.
Furthermore, solutions such as MergeShuffle \cite{Bacher_et_al_2015} can be considered for enabling parallelism within the permutation step.

Instructions in Step~2 and Step~3 blocks correspond to the evaluation of
  subsequent terms of equation~(\ref{eq:p}) and the collision event handling.
To exemplify the way the GPU and CPU backends are engineered in PySDM, the 
  implementation of the pair flag routine with ThrustRTC, and the 
  update-attributes routine with Numba are given in Fig.~\ref{fig:pair_flag} and Fig.~\ref{fig:coala}, respectively.
   
\begin{figure}[t]
\begin{lstlisting}[language=Python]
# step 0: removal of super-droplets with zero multiplicity
remove_if_equal_0(idx,                                  # in/out
                  multiplicities)                       # in

# step 1: cell-wise shuffling, pair flagging
urand(u01)
shuffle_per_cell(cell_start,                            # out
                 idx,                                   # in/out
                 cell_idx, cell_id, u01)                # in

flag_pairs(pair_flag,                                   # out
           cell_id, cell_idx, cell_start)               # in 

# step 2: collision probability evaluation
coalescence_kernel(pair_prob,                           # out
                   pair_flag, volume_view)              # in
                   
times_max(pair_prob,                                    # in/out 
          multiplicities, pair_flag)                    # in

normalize(pair_prob,                                    # in/out
          dt, dv, cell_id, cell_idx, cell_start)        # in

# step 3: collision triggering and attribute updates
urand(u01[:N_SD//2])
compute_gamma(pair_prob,                                # in/out
              u01[:N_SD//2])                            # in

update_attributes(multiplicities, attributes,           # in/out 
                  pair_prob)                            # in
\end{lstlisting}
  \caption{Algorithm workflow within a timestep,
     in/out comments mark argument intent. 
  }\label{fig:algo}
\end{figure}

\begin{figure}[t]
    \centering
\begin{lstlisting}[language=Python]
_kernel = ThrustRTC.For(
    ['perm_cell_start', 'perm_cell_id', 'pair_flag', 'length'], "i", '''
    pair_flag[i] = (
        i < length - 1 &&
        perm_cell_id[i] == perm_cell_id[i+1] &&
        (i - perm_cell_start[perm_cell_id[i]]) % 2 == 0
    );
    ''')

def flag_pairs(pair_flag, cell_start, cell_id, cell_idx):
    perm_cell_id = ThrustRTC.DVPermutation(cell_id.data, cell_id.idx.data)
    perm_cell_start = ThrustRTC.DVPermutation(cell_start.data, cell_idx.data)
    d_length = ThrustRTC.DVInt64(len(cell_id))
    _kernel.launch_n(len(cell_id), 
        [perm_cell_start, perm_cell_id, cell_idx.data, 
         pair_flag.indicator.data, d_length])
\end{lstlisting}
    \caption{GPU backend implementation of the pair-flagging routine.}
    \label{fig:pair_flag}
\end{figure}

\begin{figure}[t]
    \centering
\begin{lstlisting}[language=Python]
@numba.njit(parallel=True, error_model='numpy')
def _update_attributes(length, n, attributes, idx, gamma):
    for i in prange(length//2):
        j = idx[2*i]
        k = idx[2*i + 1]
        if n[j] < n[k]:
            j, k = k, j
        g = min(int(gamma[i]), int(n[j] / n[k]))
        if g == 0:
            continue
        new_n = n[j] - g * n[k]
        if new_n > 0:
            n[j] = new_n
            for attr in range(0, len(attributes)):
                attributes[attr, k] += g * attributes[attr, j]
        else:  # new_n == 0
            n[j] = n[k] // 2
            n[k] = n[k] - n[j]
            for attr in range(0, len(attributes)):
                attributes[attr, j] = attributes[attr, j] * g \
                                    + attributes[attr, k]
                attributes[attr, k] = attributes[attr, j]
                
def update_attributes(n, intensive, attributes, gamma):
    _update_attributes(len(n.idx),
        n.data, intensive.data, attributes.data,      # in/out
        n.idx.data, gamma.data)                       # in
\end{lstlisting}
    \caption{CPU backend implementation of the attribute-update routine.\\~\\a}\label{fig:coala}
\end{figure}

\section{Example simulations}\label{sec:examples}

This section outlines a set of test cases useful in validating
  implementation of the proposed API (and included in the set of examples shipped with PySDM).
First, a~simulation constructed analogously as that reported in Fig.~2 in
  the original SDM paper \cite{Shima_et_al_2009} is proposed.
The considered volume of $V=10^6\text{ } \text{m}^3$ is populated with $N_0\cdot V = 2^{23} \cdot 10^6$ particles. 
Sizes of the particles follow an exponential particle volume distribution $p(v_{[i]}) = \overline{v}^{-1} \exp(-v_{[i]}/\overline{v})$ where $\overline{v} = (30.531 \mu\text{m})^3 4 \pi/3$.
The simulation uses $2^{17}$ super-droplets initialised with equal multiplicities so that 
  at $t=0\,\text{s}$ each one represents a quantile of the total number of modeled particles
  (within an arbitrarily chosen percentile range from 0.001\% to 99.999\%).
The timestep is~$1\,\text{s}$.
The Golovin \cite{Golovin_1963} additive $a(v_1, v_2) = b (v_1 + v_2)$ coalescence kernel with $b=1500 \text{s}^{-1}$ is used 
  for which an analytical solution to the Smoluchowski's equation is known (cf. equation~36 in \cite{Golovin_1963}):
\begin{equation}
        \phi(v, t) =
            \frac{1 - \tau(t)}{v \sqrt{\tau(t)}} 
            {\rm I^1_{ve}}\!\!\left(\frac{2v\sqrt{\tau(t)}}{\overline{v}}\right) 
            \exp\!\left(\frac{-v(1 + \tau(t) -2  \sqrt{\tau(t)}\,)}{\overline{v}}\right)
\end{equation}
where $\tau = 1 - \exp(-N_0 b \overline{v} t)$
and ${\rm I^1_{ve}}$ stands for exponentially scaled modified Bessel function of the first kind (\code{scipy.special.ive}).
  
\begin{figure}[t]
\begin{minipage}[t]{0.49\linewidth}
    \includegraphics[width=\textwidth]{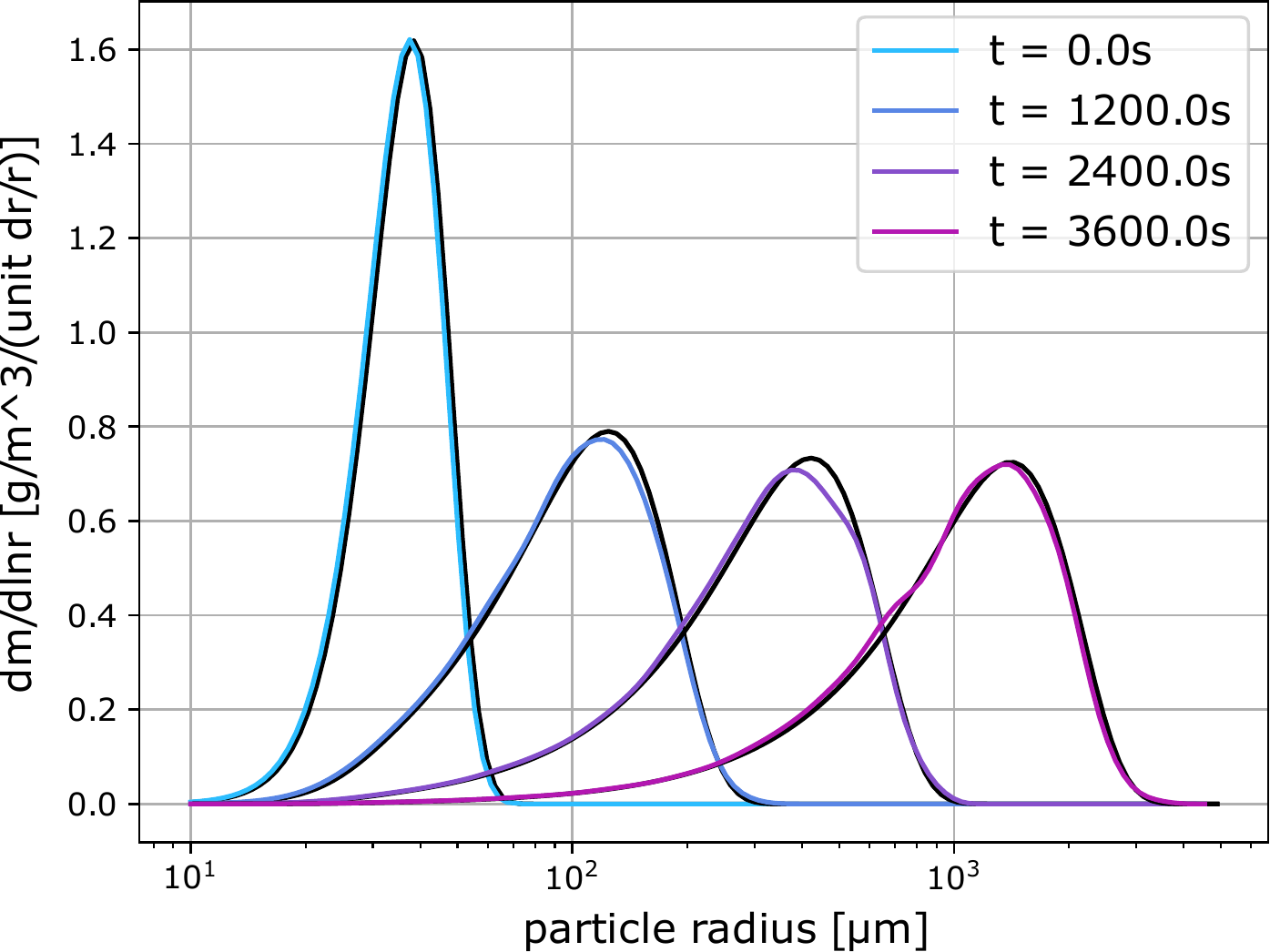}
    \caption{
    Particle mass spectrum:
    SDM results (colour) vs. analytic solution (black).
    }
    \label{fig:shima_fig_2}
\end{minipage}
\hspace{.02\linewidth}
\begin{minipage}[t]{0.49\linewidth}
  \includegraphics[width=.98\textwidth]{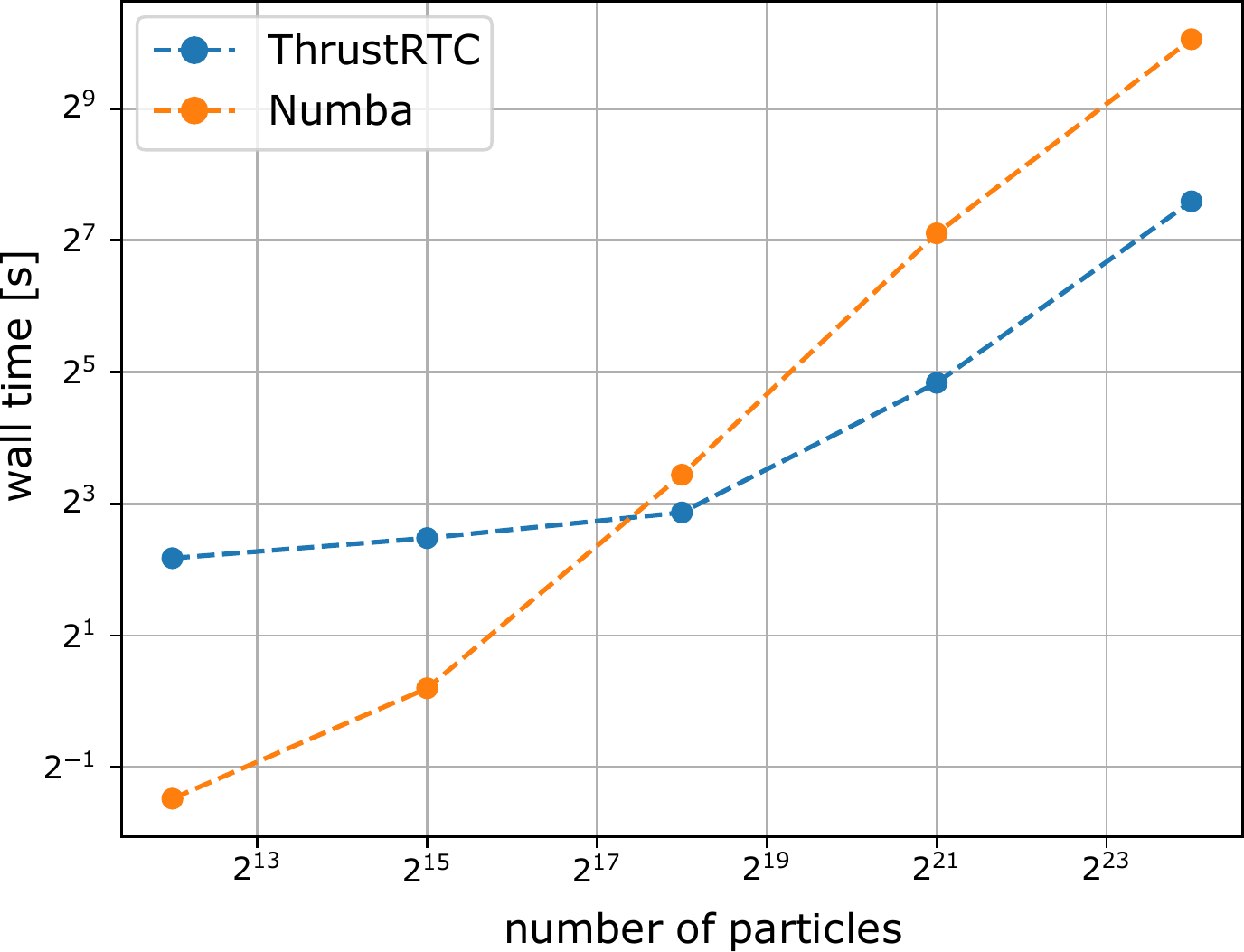}
  \caption{
  Simulation wall time as a function of the number of super particles used.
  } 
\label{fig1}
\end{minipage}
\end{figure}

Plot \ref{fig:shima_fig_2} shows the relationship between the mass of droplets per unit of $\text{ln} r$ 
  and the droplet radius $r$ (assuming particle density of $1000\,\text{kg/m\textsuperscript{3}}$ as for liquid water). 
Results obtained with {PySDM} are plotted by aggregating super-particle multiplicities onto a grid of ca. 100 bins, 
  and smoothing the result twice with a running average with a centered window spanning five bins.

Figure~\ref{fig1} documents simulation wall times for the above test case 
  measured as a function of number of super-particles employed.
Wall times for CPU (Numba) and GPU (ThrustRTC) backends are
  compared depicting a five-fold GPU-to-CPU speedup for large state vectors (tested on commodity hardware: Intel Core i7-10750H CPU and NVIDIA GeForce RTX 2070 Super Max-Q GPU).

\begin{figure}[!ht]
    \centering
    \includegraphics[width=1\textwidth]{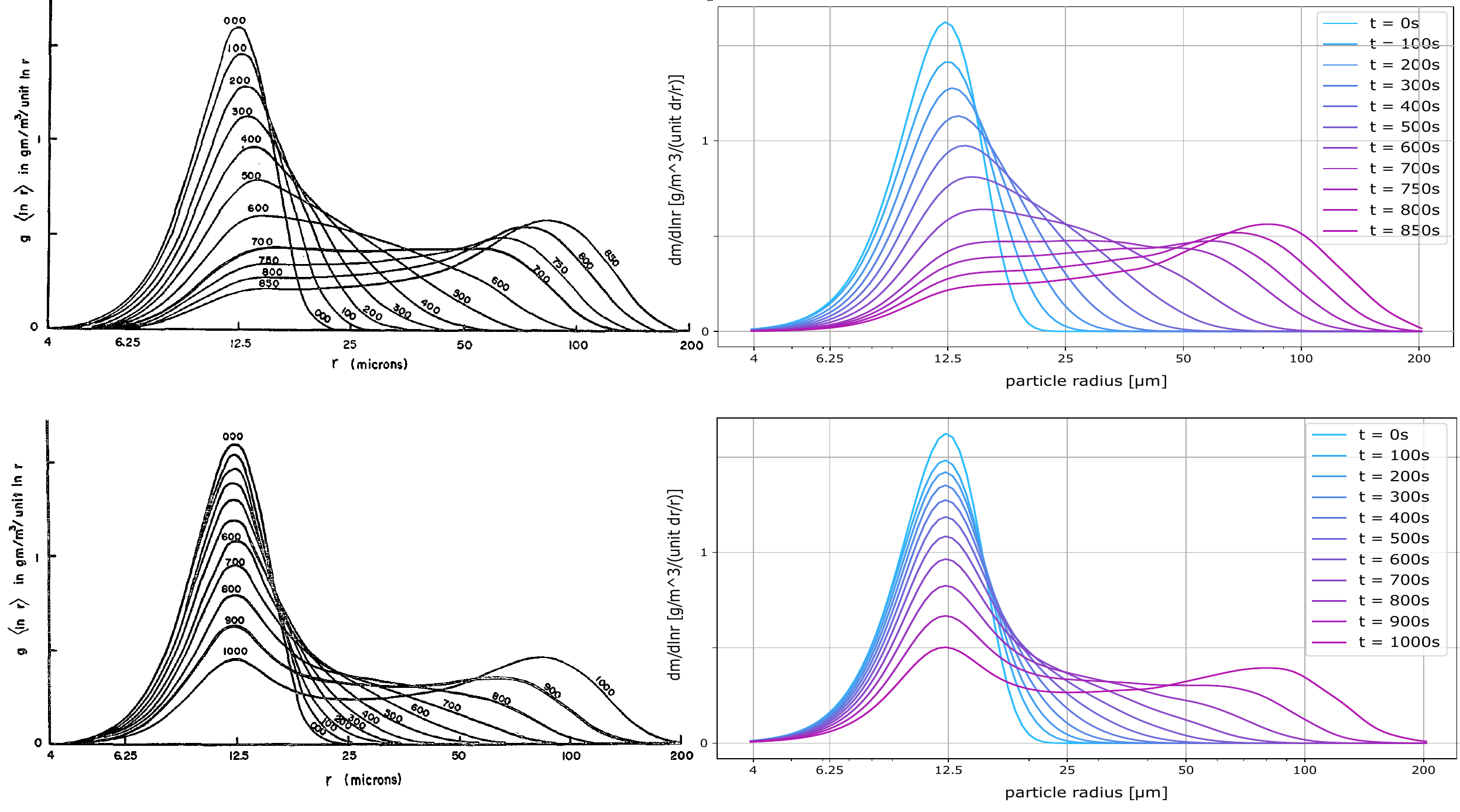}
    \caption{Left panels show Figs. 5 and 8 from \cite{Berry_1966}
    (copyright: American Meteorological Society, used with permission). 
    Right panels depict solutions obtained with PySDM.
    Top: gravitational kernel; bottom: kernel modelling electric field effect.}
    \label{fig:berry}
\end{figure}

Figure \ref{fig:berry} presents results visualised in analogous plots
  but from simulations with more physically-relevant 
  coalescence kernels.
Simulation setup follows the work of \cite{Berry_1966} and features
  so-called gravitational kernel involving a parametric form of particle
  terminal velocity dependence on its size, and a parametric kernel modeling the
  the electric field effects on collision probability.
Since the analytic solution is not known in such cases, the results are
  juxtaposed with figures reproduced from \cite{Berry_1966}.


\section{Summary and discussion}\label{sec:concl}

This paper has discussed a set of data structures and computational kernels constituting 
  a number-crunching backend API for the Super-Droplet Method Monte-Carlo algorithm
  for representing collisional growth of particles.
Employment of the API assures separation of concerns, in particular separation of
  parallelisation logic embedded in the backend-level routines from domain logic
  pertinent to the algorithm workflow.
This improves code readability, paves the way for modular design and testability,
  which all contribute to code maintainability.

The presented SDM algorithm and API descriptions discern data dependencies across
  the steps of the algorithm (including in/out parameter ``intent'') and highlight
  parallelisation opportunities in different steps. 
Most of the discerned steps of the SDM algorithm are characterised by some degree of
  freedom in terms of their implementation.
Embracing the API shall help in introducing a dependency injection mechanism allowing
  unit testing of selected steps and profiling performance with different variants
  of implementation.

The design of the API has been proved, in the sense of reused abstraction principle, 
  within the PySDM project \cite{Bartman_et_al_2021} where two backends sharing 
  the API offer contrasting implementations for CPU and GPU computations.
Both~backends are implemented in Python, however: (i) they are targeting different
  hardware (CPU vs. GPU), (ii) they are based on different underlying technology 
  (Numba: LLVM-based JIT compiler, and ThrustRTC: NVRTC-based runtime compilation mechanism for CUDA), and (iii) they even do not share the de-facto-standard Python Numpy 
  arrays as the storage layer.
This highlights that the introduced API is not bound to particular 
  implementation choices, and in principle its design is applicable 
  to other languages than Python.

It is worth pointing out that, in the super-particle CFD-coupled
  simulations context SDM was introduced and gained attention in, the scheme is particularly well suited for leveraging modern hybrid CPU-GPU hardware.
First, the algorithm is (almost) embarrassingly parallel.
Second, the CPU-GPU transfer overhead is not a bottleneck when GPU-resident dispersed phase representation (super-particles) is coupled with CPU-computed CFD for the continuous phase (fluid flow) as only statistical moments of the size spectrum of the particles are needed for the CFD coupling.
Third, the CPU and GPU resources on hybrid hardware can be leveraged effectively if 
fluid advection (CPU) and particle collisions (GPU) are computed simultaneously 
(see \cite{Dziekan_et_al_2019}).

Overall, the discussed API prioritises simplicity and was intentionally presented 
  in a paradigm-agnostic pseudo-code-mimicking way, leaving such aspects as object
  orientation up to the implementation.
Moreover, while the presented API includes data structures and algorithm steps 
  pertinent to multi-dimensional CFD grid coupling, presented examples featured
  zero-dimensional setups, for brevity.
In PySDM \cite{Bartman_et_al_2021}, the backend API is extended 
  to handle general form of coalescence kernels, representation of particle
  displacement, condensational growth of particles, aqueous chemical reactions, and the examples shipped 
  with the package include simulations in multiple physical dimensions.

\vspace{-.5em}
\subsection*{Acknowledgements}
We thank Shin-ichiro Shima and Michael Olesik for helpful discussions as well as an anonymous reviewer for insightful comments.
Thanks are due to Numba and ThrustRTC developers for support in using the packages.

\vspace{-.5em}
\bibliographystyle{splncs04}
\bibliography{bibliography}
\end{document}